\documentclass[twocolumn,showpacs,preprintnumbers,amsmath,amssymb]{revtex4}

\usepackage{graphicx}
\usepackage{dcolumn}
\usepackage{bm}
\usepackage{amsmath,amssymb}

\begin{document}

\preprint{APS/123-QED}

\title{Magnetic properties of quasi-two-dimensional $S$ = 1/2 Heisenberg antiferromagnet with distorted square lattice}

\author{Hironori Yamaguchi}
\author{Yusuke Tamekuni}
\author{Yoshiki Iwasaki}
\author{Rei Otsuka}
\author{Yuko Hosokoshi}
\affiliation{Department of Physical Science, Osaka Prefecture University, Osaka 599-8531, Japan}
\author{Takanori Kida}
\author{Masayuki Hagiwara}
\affiliation{Center for Advanced High Magnetic Field Science (AHMF), Graduate School of Science, Osaka University, Osaka 560-0043, Japan}

Second institution and/or address\\
This line break forced

\date{\today}

\begin{abstract}
We successfully synthesize single crystals of the verdazyl radical $\alpha$-2,3,5-Cl$_3$-V.
$Ab$ $initio$ molecular orbital calculations indicate that the two dominant antiferromagnetic interactions, $J_{\rm{1}}$ and $J_{\rm{2}}$ ($\alpha =J_{\rm{2}}/J_{\rm{1}}\simeq 0.56$), form an $S$ = 1/2 distorted square lattice.
We explain the magnetic properties based on the $S$ = 1/2 square lattice Heisenberg antiferromagnet using the quantum Monte Carlo method, and examine the effects of the lattice distortion and the interplane interaction contribution.
In the low-temperature regions below 6.4 K, we observe anisotropic magnetic behavior accompanied by a phase transition to a magnetically ordered state.
The electron spin resonance signals exhibit anisotropic behavior in the temperature dependence of the resonance field and the linewidth.
We explain the frequency dependence of the resonance fields in the ordered phase using a mean-field approximation with out-of-plane easy-axis anisotropy, which causes a spin-flop phase transition at approximately 0.4 T for the field perpendicular to the plane. 
Furthermore, the anisotropic dipole field provides supporting information regarding the presence of the easy-axis anisotropy.
These results demonstrate that the lattice distortion, anisotropy, and interplane interaction of this model are sufficiently small that they do not affect the intrinsic behavior of the $S$ = 1 / 2 square lattice Heisenberg antiferromagnet.
\end{abstract}


\maketitle
\section{INTRODUCTION}
One of the key focus areas of condensed matter physics is determination of the effect of quantum fluctuations on macroscopic quantum phenomena. 
Since the discovery of high-temperature superconductors in layered cuprates, considerable attention has been devoted to their parent spin system, i.e., the $S$ = 1/2 two-dimensional (2D) square lattice Heisenberg antiferromagnet (SLHAF). 
Extensive studies on the $S$ = 1/2 SLHAF have established that the ground state exhibits a long-range N$\rm{\Acute{e}}$el order at zero-temperature~\cite{square}. 
The quantum fluctuations, however, reduce the magnetic moment per site by approximately 40 ${\%}$ with respect to the classical value, and cause renormalization of the spin wave energy~\cite{square, re_spinwave, re_spinwave2}. 
In the high-field region for the $S$ = 1/2 SLHAF, it has been predicted that hybridization of the single-magnon branch with the two-magnon continuum induces instability of the single-magnon state~\cite{roton1,roton2,roton3}.  
Accordingly, the excitation spectra exhibit significant deviations from the linear spin-wave theory and form a roton-like structure, the softening of which suggests a phase transition to a modulated ground state~\cite{kubo}.

Although the long-range order (LRO) is destroyed by the quantum fluctuations in the Heisenberg case at any finite temperature~\cite{Mermin}, reduction of the spin dimensionality suppresses the quantum fluctuations and induces phase transitions at finite temperatures. 
Further, the presence of easy-axis anisotropy renders the spin system Ising-type and yields an Ising-like phase transition to the N$\rm{\Acute{e}}$el ordered state at a certain, critical temperature~\cite{Cuccoli}.
For easy-plane anisotropy, the spin system can be described as an XY-type antiferromagnet and exhibits a Berezinskii-Kosterlitz-Thouless (BKT) transition at a critical temperature $T_{\rm{BKT}}$~\cite{KT0,KT}. 
Below $T_{\rm{BKT}}$, the spin vortices form bound pairs and the spin-spin correlation function decays algebraically. 
The application of a magnetic field to the 2D Heisenberg antiferromagnet also induces the BKT transition, where spin fluctuations along the field direction are suppressed, yielding an effective easy-plane anisotropy~\cite{H_KT1,H_KT2}.
However, as real materials inevitably possess finite interplane interactions, observation of the BKT transition is difficult.
A second-order phase transition to a three-dimensional (3D) LRO occurs at a higher temperature than $T_{\rm{BKT}}$ in practice. 
Nonmonotonic field dependence of the phase transition temperature can evidence the correlation associated with the BKT transition caused by the field-enhanced easy-plane anisotropy in quasi-2D systems~\cite{H_KT2}.

Many model compounds for the $S$ = 1/2 SLHAF have been reported to date, the majority of which are based on copper oxide, such as La$_2$CuO$_4$.
Although experimental studies have revealed the fundamental properties on such compounds, the large exchange interactions of the copper oxides have rendered complete examination of the field dependence of the magnetic behavior difficult.
Accordingly, synthesis of materials with lower exchange-interaction energy scales has been implemented, so as to realize alternative model compounds. 
The molecular-based complexes (5CAP)$_2$CuX$_4$ and (5MAP)$_2$CuX$_4$, where X = Cl or Br, exhibit 2D layers composed of CuX$_4$$^{2-}$ anions~\cite{CAP1,CAP2,CAP3}.
Their relatively weak exchange interactions enable observation of the magnetic properties up to the saturation fields.
Recently, materials based on 2D arrays of magnetic Cu$^{2+}$ ions linked by organic ligands have been intensively studied ~\cite{H_KT3, good_withESR,kohama,menkan}.
In these studies, high two-dimensionality was constructed through direct overlapping of the orbitals between Cu and pyrazine, and some of the resultant materials exhibited the expected nonmonotonic behavior of the phase transition temperature~\cite{H_KT3, good_withESR,kohama}.
Furthermore, it has been reported that the two-dimensionality can be tuned through the substitution of ligand between layers~\cite{menkan}. 
These experimental studies on the Cu-based complexes have verified the quantum fluctuation effects on the $S$ = 1/2 SLHAF and stimulated investigation of the quantum nature of square-based lattices.

In this paper, we report a new model compound of the $S$ = 1/2 square-based lattice.
We successfully synthesize single crystals of the verdazyl radical $\alpha$-2,3,5-Cl$_3$-V [= 3-(2,3,5-trichlorophenyl)-1,5-diphenylverdazyl].
Then, $ab$ $initio$ molecular orbital (MO) calculations indicate that the two dominant antiferromagnetic (AF) interactions form an $S$ = 1/2 distorted square lattice.
We explain the magnetic properties based on the $S$ = 1/2 SLHAF using the quantum Monte Carlo (QMC) method and examine the effects of the lattice distortion.
In the low-temperature regions, we observe anisotropic magnetic behavior accompanied by a phase transition to an ordered state.
In addition, we explain the frequency dependence of the electron spin resonance (ESR) fields in the ordered phase using a mean-field approximation with out-of-plane easy-axis anisotropy.
The anisotropic energy derived from the dipole-dipole interactions provides supporting information regarding the presence of the easy-axis anisotropy.
Through analysis of the experimental results, we evaluate the exchange and anisotropy constants and confirm that $\alpha$-2,3,5-Cl$_3$-V is a new model compound with an $S$ = 1/2 square-based lattice.

\section{EXPERIMENTAL}
Synthesis of 2,3,5-Cl$_3$-V, the molecular structure of which is shown in Fig. 1(a), was performed using a procedure similar to that used to prepare the typical verdazyl radical 1,3,5-triphenylverdazy~\cite{synthesis}.  
Recrystallization in acetonitrile yielded deep-green crystals of $\alpha$ (block) and $\beta$ (needle)
phases, which are defined based on the unit cell volume per number of molecules ($V/Z$) at room temperature (RT).
X-ray intensity data were collected using a Rigaku AFC-7R mercury CCD diffractometer and a Rigaku AFC-8R mercury CCD RA-micro7 diffractometer at RT and 25 K, respectively, with graphite-monochromated Mo K$\rm{\alpha}$ radiation and a Japan Thermal Engineering cryogenic He gas flow (XR-HR10K). 
The structure was determined via a direct method using the SIR2004~\cite{SIR2004} and was refined using the SHELXL97 crystal structure refinement program~\cite{SHELX-97}.
The structural refinement was carried out using anisotropic and isotropic thermal parameters for the nonhydrogen and the hydrogen atoms, respectively. 
All the hydrogen atoms were placed at the calculated ideal positions. 

The magnetic susceptibility and magnetization curves were measured using a commercial SQUID magnetometer (MPMS-XL, Quantum Design).
High-field magnetization measurement in pulsed magnetic fields of up to approximately 50 T was conducted using a non-destructive pulse magnet at the Center for Advanced High Magnetic Field Science (AHMF), Osaka University.
The experimental results were corrected for the diamagnetic contribution ($-3.20{\times}10^{-4}$ emu mol$^{-1}$), which was determined based on the QMC analysis and close to the value calculated using the Pascal method.
The specific heat was measured with a commercial calorimeter (PPMS, Quantum Design) using a thermal relaxation method.
The ESR measurements were performed utilizing a vector network analyzer (ABmm) and a superconducting magnet (Oxford
Instruments) at AHMF, Osaka University.
At approximately 19.6 GHz, we used a laboratory-built cylindrical high-sensitivity cavity. 
X-band (9.47 GHz) ESR measurements were conducted using a Bruker EMX Plus spectrometer at the Institute for Molecular Science.
All experiments were performed using single crystals with typical dimensions of 2.0$\rm{\times}$0.5$\rm{\times}$0.5 mm$^3$ 

$Ab$ $initio$ MO calculations were performed using the UB3LYP method as broken-symmetry (BS) hybrid density
functional theory calculations. 
All calculations were performed using the GAUSSIAN09 software package and 6-31G basis sets. 
The convergence criterion was set to 10$^{-8}$ hartree.
To estimate the intermolecular exchange interaction of the molecular pairs within 4.0 $\rm{\AA}$ , we employed a conventional evaluation scheme~\cite{MOcal}. 

The QMC code utilized in this study was based on the directed loop algorithm in the stochastic series expansion representation~\cite{QMC1}. 
The calculation was performed on finite-size lattices with a linear size up to $L$ = 32, which involve under a periodic boundary condition in total $N$ = 4$L^2$ spins, and it is confirmed that there is no size dependence.
All calculations were conducted using the Algorithms and Libraries for Physics Simulations (ALPS) application~\cite{ALPS1,ALPS2,ALPS3}.

\section{RESULTS}
\subsection{Crystal structure and magnetic model}
The crystallographic data for the synthesized $\alpha$-2,3,5-Cl$_3$-V are summarized in Table I~\cite{CCDC}, and the molecular structure is shown in Fig. 1(a).
The verdazyl ring (which includes four N atoms), the upper two phenyl rings, and the bottom 2,3,5-trichlorophenyl ring are labeled ${\rm{R}_{1}}$, ${\rm{R}_{2}}$, ${\rm{R}_{3}}$, and ${\rm{R}_{4}}$, respectively.
The molecule is no longer planar owing to electrostatic repulsion between the Cl and N atoms, and the dihedral angles of ${\rm{R}_{1}}$-${\rm{R}_{2}}$, ${\rm{R}_{1}}$-${\rm{R}_{3}}$, ${\rm{R}_{1}}$-${\rm{R}_{4}}$ are approximately 36$^{\circ}$, 22$^{\circ}$, and 82$^{\circ}$, respectively.
The results of the MO calculations indicate that approximately 66 ${\%}$ of the total spin density is present on ${\rm{R}_{1}}$. 
Further, while ${\rm{R}_{2}}$ and ${\rm{R}_{3}}$ each account for approximately 15 ${\%}$ and 17 ${\%}$ of the relatively large total spin density, ${\rm{R}_{4}}$ accounts for less than 2 ${\%}$ of the total spin density.
The extremely small value of the spin density on ${\rm{R}_{4}}$ originates from the discontinuity of the $\pi$-orbitals, owing to the large dihedral angle of 82$^{\circ}$.
Therefore, the intermolecular interactions are caused by the short contacts of N or C related to the ${\rm{R}_{1}}-{\rm{R}_{3}}$ rings.
Note that, because this study focuses on the low-temperature magnetic properties, the crystallographic data obtained at 25 K are used hereafter. 

We performed $ab$ $initio$ MO calculations to quantitatively evaluate the dominant intermolecular magnetic interactions on all molecular pairs within 4.0 $\rm{\AA}$. 
Consequently, we found three types of AF interactions, i.e., $J_{\rm{1}}$, $J_{\rm{2}}$, and $J_{\rm{3}}$.
These interactions were evaluated to be $J_{\rm{1}}/k_{\rm{B}}$ = 14.8 K, $J_{\rm{2}}/k_{\rm{B}}$ = 8.3 K ($\alpha =J_{\rm{2}}/J_{\rm{1}}\simeq 0.56$), and $J_{\rm{3}}/k_{\rm{B}}$ = 3.5 K ($\beta  =J_{\rm{3}}/J_{\rm{1}}\simeq 0.24$), where $k_{\rm{B}}$ is the Boltzmann constant, which are defined in the following eq. (1).
The molecular pairs related to $J_{\rm{1}}$ and $J_{\rm{2}}$ have C-N and C-C short contacts of 3.39 and 3.33 $\rm{\AA}$ $[$Figs. 1(b) and 1(c)$]$, respectively, both of which are related by a two-fold screw axis parallel to the $b$-axis.
The molecular pair related to $J_{\rm{3}}$ has a C-C short contact of 3.71 $\rm{\AA}$ $[$Fig. 1(d)$]$, which is doubled by an inversion symmetry.
The two dominant AF interactions, i.e., $J_{\rm{1}}$ and $J_{\rm{2}}$, form an $S$ = 1/2 square lattice in the $ab$-plane, as shown in Fig 1(e), and those 2D planes are partially connected by the weak AF interaction $J_{\rm{3}}$, as shown in Figs. 1(f).
The corresponding 2D square lattice and 3D stacking of the planes are shown in Figs. 1(g) and 1(h), respectively.  
As the interplane coordination number is smaller than that for simple cubic stacking, two-dimensionality should be enhanced in the present lattice.
The lattice distortion in the 2D plane, which is characterized by the value of $\alpha$, can be described as chain-type and induces an intermediate state between the one-dimensional (1D) chain and the 2D square lattice. 
Very recently, theoretical and experimental studies of such lattice system with $\alpha$ = 0.35 is reported in the Cu-based compound Cu(en)(H$_2$O)$_2$SO$_4$ (CUEN)~\cite{zigzag}.
We discuss our results comparing with those of the reported compound.

\begin{table}
\caption{Crystallographic data for $\alpha$-2,3,5-Cl$_3$-V}
\label{t1}
\begin{center}
\begin{tabular}{ccc}
\hline
\hline 
Formula & \multicolumn{2}{c}{C$_{20}$H$_{14}$Cl$_{3}$N$_{4}$}\\
Crystal system & \multicolumn{2}{c}{Monoclinic}\\
Space group & \multicolumn{2}{c}{$P$2$_1$/$n$}\\
Temperature (K) & RT & 25(2)\\
Wavelength ($\rm{\AA}$) & \multicolumn{2}{c}{0.7107} \\
$a (\rm{\AA}$) &  9.009(3) &  8.925(3) \\
$b (\rm{\AA}$) &  10.648(3) & 10.567(4) \\
$c (\rm{\AA}$) &  19.838(6) & 19.428(7)\\
$\beta$ (degrees) &  101.821(4) & 101.624(5)\\
$V$ ($\rm{\AA}^3$) & 1862.7(10) &  1794.7(11) \\
$Z$ & \multicolumn{2}{c}{4} \\
$D_{\rm{calc}}$ (g cm$^{-3}$) & 1.486 & 1.542\\
Total reflections & 3248 & 2909\\
Reflection used & 2840 & 2716\\
Parameters refined & \multicolumn{2}{c}{244}\\
$R$ [$I>2\sigma(I)$] & 0.0373 & 0.0305\\
$R_w$ [$I>2\sigma(I)$] & 0.0946 & 0.0854\\
Goodness of fit & 1.045 & 0.951\\
CCDC &  1530390 & 1530389 \\
\hline
\hline
\end{tabular}
\end{center}
\end{table}

\begin{figure*}[t]
\begin{center}
\includegraphics[width=38pc]{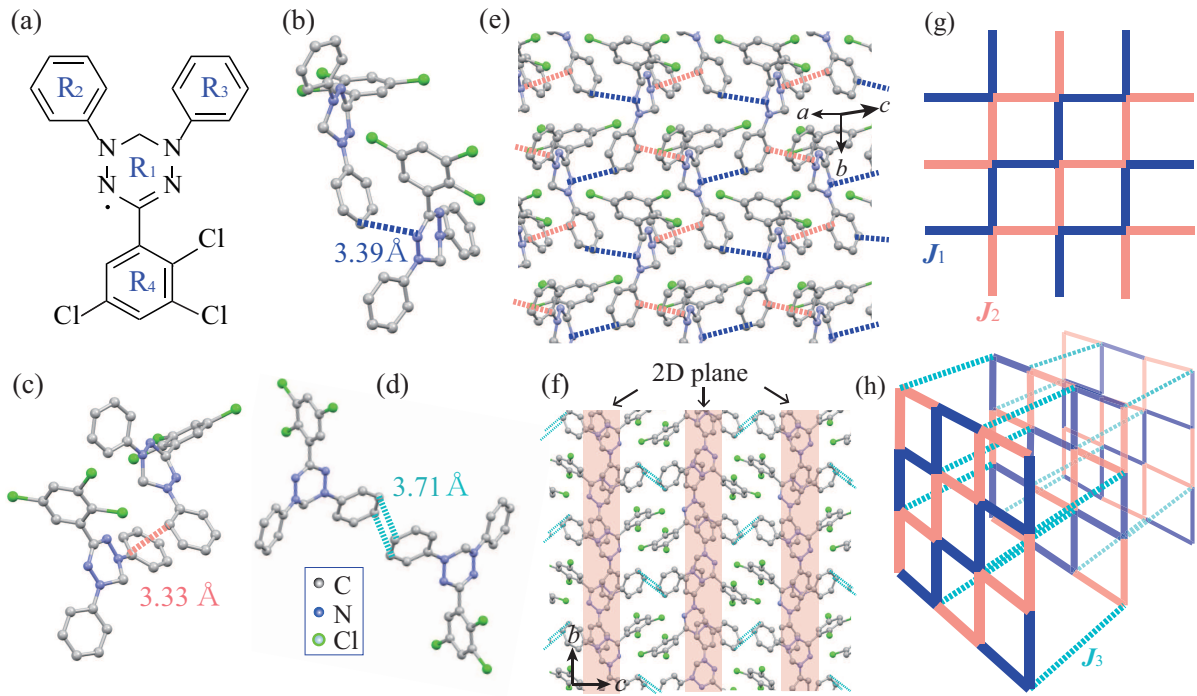}
\caption{(color online) (a) Molecular structure of 2,3,5-Cl$_3$-V. Molecular pairs associated with the magnetic interactions (b) $J_{\rm{1}}$, (c) $J_{\rm{2}}$, (d) $J_{\rm{3}}$. The broken lines indicate C-N and C-C short contacts. Crystal structure of $\alpha$-2,3,5-Cl$_3$-V viewed (e) almost perpendicular and (f) parallel to the $a$-axis. Hydrogen atoms are omitted for clarity. (g) Square lattice in the $ab$-plane formed by $J_{\rm{1}}$ and $J_{\rm{2}}$ and (h) 3D lattice composed of the 2D planes partially connected by interplane interaction $J_{\rm{3}}$.}\label{f1}
\end{center}
\end{figure*}

\subsection{Magnetic and thermodynamic properties}
Figure 2 shows the temperature dependence of the magnetic susceptibility ($\chi$ = $M/H$) for $B{\perp}ab$ and $B{\parallel}ab$ at 0.2 T.
We observed a broad peak at approximately 12 K for both field directions, which indicates an AF short-range order (SRO) in the 2D square lattice.
Below 6.4 K, a significant difference between $B{\perp}ab$ and $B{\parallel}ab$ appears, which indicates that a phase transition to a 3D LRO occurs at $T_N$ = 6.4 K, with the aid of the interplane coupling $J_{\rm{3}}$.
For $T$ ${\textgreater}$ $T_N$, there is no significant difference between the field directions owing to the isotropic nature of organic radical systems.
This finding is also supported by isotropic $g$-values obtained via ESR measurements, which are described below.
The Weiss temperature is estimated to be ${\theta}_{\rm{W}}$ = -14(1) K above 100 K.
Because the anisotropic magnetic behavior appears only below $T_N$, the energy scale of the anisotropy should be quite small.
A dipole field accompanied by the phase transition to the LRO and/or a small spin-orbit coupling in organic systems~\cite{radical_SO} is considered as a possible cause of the anisotropic behavior.

The experimental result for the specific heat $C_{\rm{p}}$ at zero-field clearly exhibits a $\lambda$-type sharp peak at $T_N$, which is associated with the phase transition to the LRO, as shown in Fig. 3(a).
The lattice contribution in the low-temperature region can be approximated as $C_{\rm{l}}=a_{1}T^{3}+a_{2}T^{5}+a_{3}T^{7}$.
The magnetic specific heat $C_{\rm{m}}$ is obtained by subtracting $C_{\rm{l}}$ with the constants $a_{1}=0.0087$, $a_{2}=-2.0\times10^{-5}$, and $a_{3}=1.9\times10^{-8}$, which are determined to fit the following calculated result by using the QMC method, as shown in Fig. 3(b). 
The evaluated lattice contribution is similar to those for some copper complexes forming the 2D SLHAF~\cite{Clattice}, and the value of $a_{1}$, which corresponds to Debye temperature of 61 K, is quite close to those for other verdazyl radical compounds~\cite{2Cl6FV, 3Cl4FV, pBrV}.
The shape of the obtained $C_{\rm{m}}$ demonstrates that the quasi-2D character of the present system yields a separation of the 3D ordering peak from the rounded behavior arising from the 2D SRO~\cite{H_KT3, C_plane}.
The magnetic entropy $S_{\rm{m}}$, which can be obtained through integration of $C_{\rm{m}}$/$T$, demonstrates that approximately 1/3 of the total magnetic entropy of 5.76 ($R$ln2) is associated with the phase transition at $T_{\rm{N}}$, while the residual entropy is almost consumed above $T_{\rm{N}}$, as shown in Fig. 3(c).
This behavior also indicates that the present system has a quasi-2D character, yielding a sufficient development of the SRO in the 2D square plane above $T_{\rm{N}}$. 
Furthermore, application of magnetic fields induces an increase in $T_{\rm{N}}$, as shown in Fig. 3(d).
Such field dependence of $T_{\rm{N}}$ is predicted for the quasi-2D Heisenberg AF system as a consequence of the field-enhanced effective easy-plane anisotropy~\cite{H_KT1,H_KT2}.
This behavior has actually been observed in some model compounds for the $S$=1/2 SLHAF~\cite{H_KT3, good_withESR,kohama}.
CUEN, which has the same lattice distortion as the present compound, also exhibits a nonmonotonic development of transition temperature~\cite{zigzag}.
Thus, the field dependence of the specific heat also evidences the quasi-2D character for the present system.
No significant difference in $C_{\rm{p}}$ was found for the results obtained for $B{\perp}ab$ and $B{\parallel}ab$.

Figure 4 and its inset show the magnetization curves in a pulsed field at 1.4 K and in a static field at 1.8 K, respectively.
We observed a distinct spin-flop phase transition at $B_{c}\approx$ 0.4 T for $B{\perp}ab$, whereas the magnetization curve for $B{\parallel}ab$ apparently exhibits a monotonic increase, as shown in the inset of Fig.4.
Such behavior is consistent with the temperature dependence of $\chi$ below $T_{\rm{N}}$ and indicates the presence of weak easy-axis anisotropy perpendicular to the $ab$-plane, which stabilizes an out-of-plane spin structure.
Above $B_{c}$, the magnetization curve for $B{\perp}ab$ approaches that for $B{\parallel}ab$, and these curves eventually become almost identical.
The high-field magnetization curve exhibits a slightly nonlinear increase originating from the 2D quantum fluctuations, as shown in Fig. 4. 
Given the isotropic $g$-value of $\sim2.00$, the saturation value of 0.96 $\mu_{\rm{B}}$/f.u. indicates that the radical purity is approximately 96 ${\%}$.
This small amount of impurity originate from unoxidized molecules and has no effect on the intrinsic magnetic properties owing to its nonmagnetic nature~\cite{3ladders,okabe}.
We consider this purity in the following analysis. 

\begin{figure}[t]
\begin{center}
\includegraphics[width=17pc]{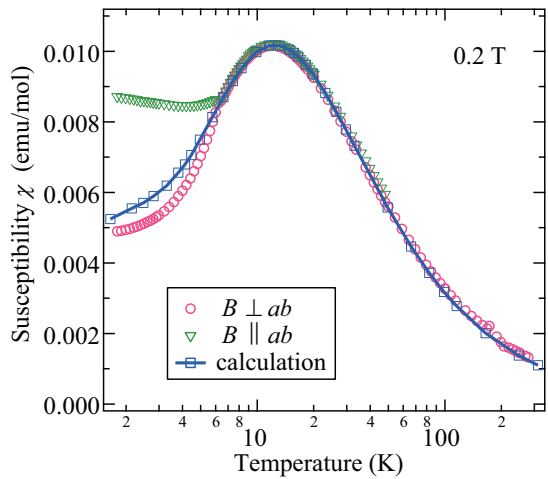}
\caption{(color online) Temperature dependence of magnetic susceptibility ($\chi$ = $M/H$) of $\alpha$-2,3,5-Cl$_3$-V for $B{\perp}ab$ and $B{\parallel}ab$ at 0.2 T obtained in field-cooling regime. The solid line with squares represents the calculated result for the $S$=1/2 SLHAF with $\alpha=J_{\rm{2}}/J_{\rm{1}}=0.56$.}\label{f2}
\end{center}
\end{figure}

\begin{figure}[t]
\begin{center}
\includegraphics[width=20pc]{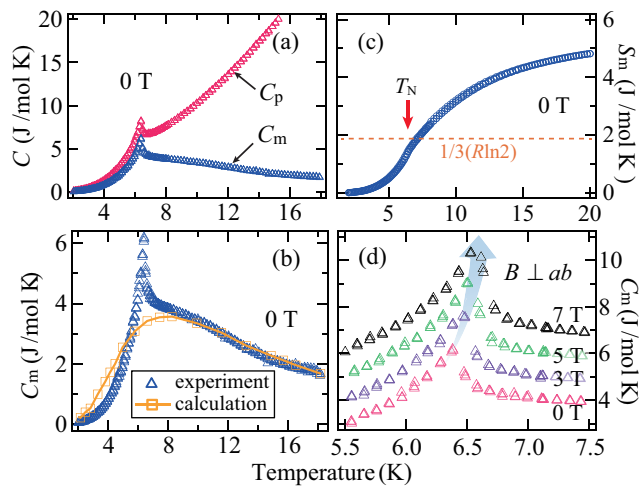}
\caption{(color online) Specific heat of $\alpha$-2,3,5-Cl$_3$-V. (a) Total specific heat $C_{\rm{p}}$ and its magnetic contribution $C_{\rm{m}}$ at 0 T. (b) Expansion of $C_{\rm{m}}$ at 0 T. The solid line with squares represents the calculated result for the $S$=1/2 SLHAF with $\alpha=J_{\rm{2}}/J_{\rm{1}}=0.56$. (c) Magnetic entropy at 0 T. (d) Expansion near the phase transition temperature at various magnetic fields for $B{\perp}ab$. For clarity, $C_{\rm{m}}$ for 3, 5, and 7 T have been shifted up by 1.0, 2.0, 3.0 J/mol K, respectively.}\label{f2}
\end{center}
\end{figure}

\begin{figure}[t]
\begin{center}
\includegraphics[width=17pc]{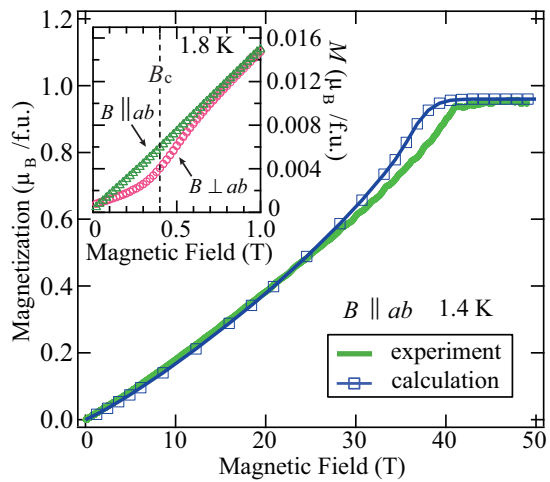}
\caption{(color online) Magnetization curve of $\alpha$-2,3,5-Cl$_3$-V at 1.4 K for $B{\parallel}ab$. The solid line with squares represents the calculated result for the $S$=1/2 SLHAF with $\alpha=J_{\rm{2}}/J_{\rm{1}}=0.56$. The inset shows magnetization curves in the low-field region below 1 T at 1.8 K for $B{\perp}ab$ and $B{\parallel}ab$. The vertical broken line indicates the spin-flop phase transition at $B_{c}$ for $B{\perp}ab$.}\label{f4}
\end{center}
\end{figure}

\subsection{Electron spin resonance}
Figures 5(a) and 5(b) show the temperature dependence of the ESR absorption spectra for $B{\perp}ab$ and $B{\parallel}ab$, respectively. 
Sharp resonance signals characteristic of organic radical systems are observed, and the resonance spectra vary dramatically in the vicinity of $T_{\rm{N}}$ for both field directions.
The temperature dependences of both the resonance field and the absorption linewidth are shown in Figs. 6(a) and 6(b), respectively. 
The resonance fields at 60 K were converted into $g$-values, yielding $g_{\perp}$=2.0037 ($B{\perp}ab$) and $g_{\parallel}$=2.0039 ($B{\parallel}ab$); these values are almost temperature-independent down to $T_{\rm{N}}$ and confirm the isotropic nature of the present system.
Below $T_{\rm{N}}$, the resonance fields rapidly shift in opposite directions, reflecting the local-field anisotropy accompanied by the phase transition.
The absorption linewidth increases rapidly below $T_{\rm{N}}$ for both directions.
Although the linewidth values for $T{\textgreater}T_{\rm{N}} $ are too small to facilitate detection of general 2D critical broadening toward the LRO~\cite{linewidth_book, ESR_2Dlinewidth}, the rapid increases in the vicinity of $T_{\rm{N}}$ are certainly related to this critical behavior. 
Assuming a conventional second-order phase transition, the ESR linewidth is expected to decrease rapidly with decreasing temperature below the transition temperature, reflecting the development of uniform internal fields, which yields a sharp peak in the linewidth~\cite{good_withESR,linewidth_exp,kimuraABX}.
While the obtained result for $B{\parallel}ab$ indeed exhibits a peak structure close to $T_{\rm{N}}$, the linewidth values remain large even at sufficiently low temperatures (below $T_{\rm{N}}$) for both field directions, as shown in Fig. 6(b). 
Additionally, a relatively large difference appears between the results for $B{\perp}ab$ and $B{\parallel}ab$ below $T_{\rm{N}}$.
The large ESR linewidth indicates inhomogeneous internal fields, which suggests a contribution of magnetic domains.
The incidence of these domains is strongly associated with the anisotropy energy in the plane normal to the applied field direction. 
The significant difference in the linewidth below $T_{\rm{N}}$ for the two field directions suggests that the in-plane anisotropy is significantly weaker than the out-of-plane easy-axis anisotropy, both of which can be derived from the dipole field as described in the following calculations.

The frequency dependence of the ESR absorption spectra in the ordered phase is presented in Figs. 7(a)-7(e).
All the resonance fields are plotted in the frequency-field diagram, as shown in Fig. 8.
The resonance signals at high frequencies in Figs. 7(a) and 7(b) are almost proportional to the external field.
Conversely, those at low frequencies in Figs. 7(c)-7(e) obviously deviate from the linear function of the field and exhibit remarkable changes across the phase transition temperature. 
At 19.56 GHz for $B{\perp}ab$ $[$Fig. 7(c)$]$, two sharp signals appear, which are emphasized by arrows. 
We regard these signals as intrinsic resonances associated with the same resonance modes as those at other frequencies. 
Because the satellite peaks indicated by the crosses have anomalously broadened shapes, they may have origins in inhomogeneous internal fields caused by a coexistent state of two phases across the first-order phase transition at $B_{\rm{c}}$. 
Note that uncertain ESR resonance signals in connection with a discontinuity of the first-order phase transition are actually observed in some AF compounds~\cite{kimuraABX, RbMn, shinozaki}.
The resonance signals indicate a zero-field gap of approximately 11 GHz, the energy scale of which corresponds to that of $B_{c}$. 
Such behavior is reminiscent of conventional AF resonance modes with easy-axis anisotropy in a two-sublattice model~\cite{kittel, MnF2}.

\begin{figure}[t]
\begin{center}
\includegraphics[width=19pc]{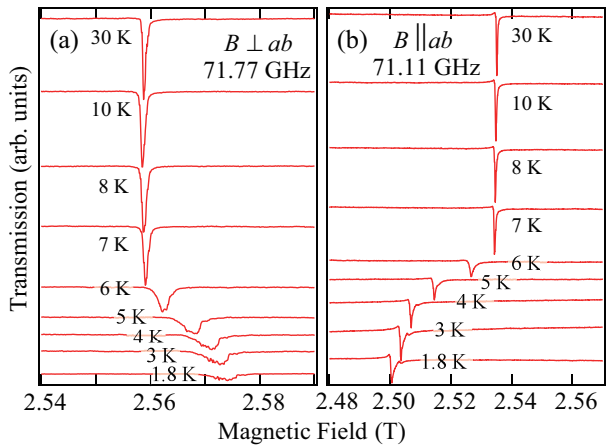}
\caption{(color online) Temperature dependence of ESR absorption spectra of $\alpha$-2,3,5-Cl$_3$-V at (a) 71.77 GHz for $B{\perp}ab$ and (b) 71.11 GHz for $B{\parallel}ab$.}\label{f4}
\end{center}
\end{figure}

\begin{figure}[t]
\begin{center}
\includegraphics[width=17pc]{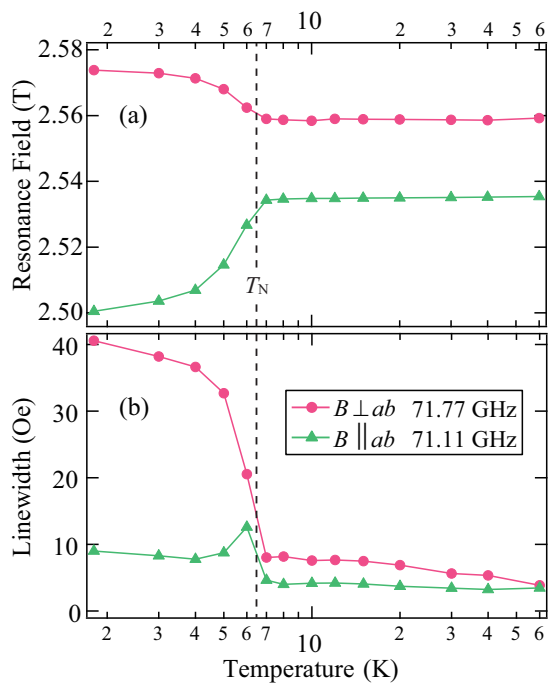}
\caption{(color online) Temperature dependence of (a) resonance field and (b) linewidht (full width at half maximum) obtained from the ESR data at 71.77 GHz for $B{\perp}ab$ and at 71.11 GHz for $B{\parallel}ab$. The solid lines are guides for the eye. The vertical broken line indicates the phase transition temperature.}\label{f4}
\end{center}
\end{figure}

\begin{figure}[t]
\begin{center}
\includegraphics[width=17pc]{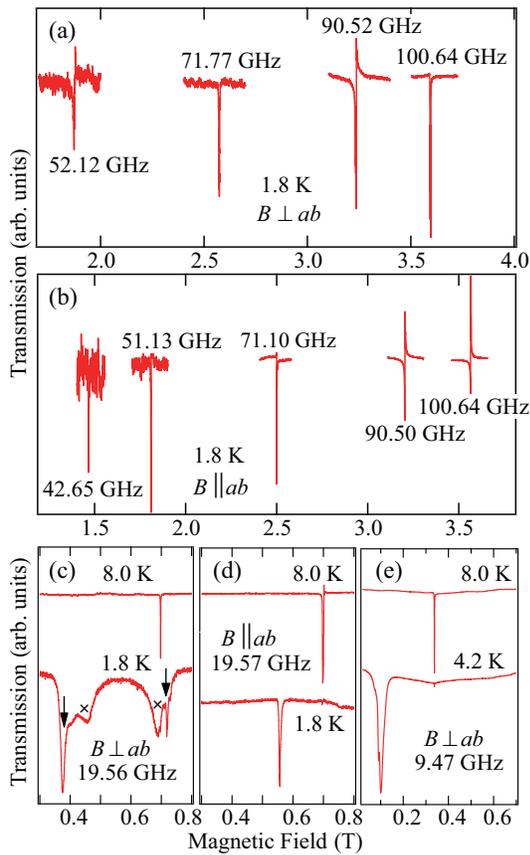}
\caption{(color online) Frequency dependence of ESR absorption spectra of $\alpha$-2,3,5-Cl$_3$-V at 1.8 K for (a) $B{\perp}ab$ and (b) $B{\parallel}ab$. ESR absorption spectra at 1.8 K and 8.0 K at (c) 19.56 GHz for $B{\perp}ab$ and (d) 19.57 GHz for $B{\parallel}ab$. The arrows and crosses in the spectrum at 1.8 K for $B{\perp}ab$ indicate intrinsic sharp signals and anomalously broadened signals, respectively. (e) ESR absorption spectra at 4.2 K and 8.0 K at 9.47 GHz (X-band) for $B{\perp}ab$.}\label{f4}
\end{center}
\end{figure}

\section{Analysis and Discussion}
\subsection{Temperature and field dependence of magnetization}
The MO calculation results indicate that the two dominant AF interactions, $J_{\rm{1}}$ and $J_{\rm{2}}$, form $S$ = 1/2 square lattices partially connected by the weak AF $J_{\rm{3}}$, as shown in Figs. 1(g) and 1(h).
Despite the fact that the phase transition to the 3D LRO occurs, the experimental results of the magnetizations and specific heat reveal that the two-dimensionality is enhanced in this spin model.
Accordingly, we analyzed the magnetic properties in terms of the $S$ = 1/2 2D system.
We calculated the magnetic susceptibility and the magnetization curve based on the $S$ = 1/2 SLHAF using the QMC method. 
The spin Hamiltonian is expressed as
\begin{equation}
\mathcal {H} = J_{\rm{1}}{\sum^{}_{<ij>}}\textbf{{\textit S}}_i{\cdot}\textbf{{\textit 
S}}_{j}+J_{\rm{2}}{\sum^{}_{<kl>}}\textbf{{\textit S}}_k{\cdot}\textbf{{\textit 
S}}_{l}-g{\mu _B}B{\sum^{}_{i}}\textbf{{\textit S}}_{i},
\end{equation}
where $\textbf{{\textit S}}$ is the spin operator, $g$ is the $g$ factor, $g$=2.00, and ${\mu}$$_B$ is the Bohr magneton. 
The MO calculations incorporating the crystallographic data at 25 K show that the exchange interactions have the relation $\alpha =J_{\rm{2}}/J_{\rm{1}}\simeq 0.56$.
Assuming this relation, good agreement was obtained between the experimental and calculated results using the parameters $J_{\rm{1}}/k_{\rm{B}}$ = 16.5 K and  $J_{\rm{2}}/k_{\rm{B}}$ = 9.2 K ($\alpha= 0.56$), as shown in Figs. 2-4.
The absolute values of the obtained parameters are quite consistent with those evaluated from the MO calculation.
A deviation appears for $\chi$ below $T_{\rm{N}}$, which arises mainly from the easy-axis anisotropy accompanied by the phase transition, as shown in Fig. 2.
For $C_{\rm{m}}$, although a large deviation accompanied by the 3D LRO appears near $T_{\rm{N}}$, the rounded behavior at approximately 9 K is well reproduced by the calculated result, as shown in Fig. 3(b)
The experimental magnetization curve also exhibits a slight deviation from the calculated result in the vicinity of the saturation field, as shown in Fig. 4. 
Since the small easy-axis anisotropy does not have a large effect on the magnetization curves at sufficiently higher fields than $B_{\rm{c}}$, the observed deviation near the saturation field is expected to be a $J_{\rm{3}}$ contribution.
The whole results exhibit relatively large deviations from the experimental data compared to those in CUEN with $\alpha= 0.35$~\cite{zigzag}. 
The deviations associated with the phase transition to the 3D LRO should be enhanced in the present compound because the interplane interaction is expected to be stronger than that for CUEN as discussed below.

We examined the $\alpha$ dependence of the magnetic behavior.
Figures 9(a)-9(c) show the $\chi$, $C_{\rm{m}}$, and low-temperature magnetization curves for representative values of $\alpha$ between 0 and 1, respectively. 
The lattice systems in the extreme cases, i.e., for $\alpha=0$ and 1, correspond to a 1D chain and a uniform 2D square lattice, respectively.
It is confirmed that the calculated results are consistent with those in Ref. ~\cite{zigzag} with different values of $\alpha$.
The value of $J_{\rm{1}}$ was determined for each $\alpha$, so as to reproduce the experimentally observed broad peak of $\chi$ associated with the SRO in the 2D plane.
Hence, the following values were obtained: $J_{\rm{1}}/k_{\rm{B}}$ = 20.7 K ($\alpha=0$), 20.1 K ($\alpha=0.1$), 18.6 K ($\alpha=0.3$),  17.0 K ($\alpha=0.5$), 15.4 K ($\alpha=0.7$), and 13.2 K ($\alpha=1$).
Although there is no distinct $\alpha$ dependence of $\chi$ in the high-temperature region, clear differences are apparent below the broad peak temperature, as shown in Fig. 9(a). 
The value of $\chi$ tends to decrease with increasing $\alpha$, and the $\alpha$ dependence becomes almost indistinguishable for $0.5<\alpha<1.0$.
These results indicate that the two-dimensionality is adequately enhanced to form a ground state characteristic of 2D spin systems for $\alpha>0.5$. 
Note that we cannot compare the low-temperature behavior of $\chi$ with the experimental result, because the phase transition to the 3D LRO occurs, along with the accompanying easy-axis anisotropy at approximately 6.4 K.
The $C_{\rm{m}}$ and magnetization curves at 0.5 K, which were calculated using parameters determined from the analysis of $\chi$, exhibit clear $\alpha$ dependence in their rounded peak and nonlinearity, respectively, as shown in Fig. 9(b) and 9(c).
With increasing $\alpha$, the rounded peak of $C_{\rm{m}}$ shows a higher value and shifts to lower temperature. 
The nonlinear increase of the magnetization curves reflects the strengths of the quantum fluctuations attributed to the low-dimensionality of the lattice system.
Further, this nonlinear increase is actually enhanced as $\alpha$ approaches 0 (i.e., towards the 1D AF chain). 
It is confirmed that the $\alpha$ dependence of both $C_{\rm{m}}$ and magnetization curve become almost indistinguishable for $0.5<\alpha<1$ as is the case with $\chi$.
These results suggest that the small chain-like lattice distortion for $0.5<\alpha<1$ does not affect the intrinsic properties of the $S$ = 1/2 SLHAF.
If we assume small $\alpha$, the experimental results deviates significantly from the calculated ones.
Therefore, the present lattice system is expected to have a distortion for $0.5<\alpha<1$, which is consistent with the evaluation from the MO calculation ($\alpha$ = 0.56).

Although QMC calculation on the 3D lattice system is beyond the scope of the present work, we can roughly evaluate the interplane interaction from the $T_{\rm{N}}$ and intraplane interactions.
The interplane interaction $J^{\rm{\prime}}$ for an $S$ = 1/2 square lattice with simple cubic stacking has a relation given by $T_{\rm{N}}=0.732\pi(J/k_{\rm{B}})/[2.43-{\rm{ln}}(J^{\rm{\prime}}/J)]$~\cite{todo}, where $J$ is a uniform interaction of the square lattice. 
If we consider $J=(J_{\rm{1}}+J_{\rm{2}})/2$, the mean field on the present distorted lattice becomes identical to that on the uniform square lattice.
Hence, using $J_{\rm{1}}/k_{\rm{B}}$ = 16.5 K, $J_{\rm{2}}/k_{\rm{B}}$ = 9.2 K, and $T_{\rm{N}}$=6.4 at zero-field, we obtain $J^{\rm{\prime}}/k_{\rm{B}}=1.4$ K. 
The interplane coordination number is 1 in the present lattice $[$Fig. 1(g)$]$, while the cubic stacking has coordination number 2.
Considering the difference in the mean field in the 3D stackings, $J_{\rm{3}}$ corresponds to $2J^{\rm{\prime}}=2.8$ K, which is almost consistent with the evaluation obtained via the MO calculation. 
The evaluated interplane interaction is stronger than that in CUEN, which causes the difference in the two-dimensionality between the two compounds.
Accordingly, it is expected that some qualitative differences in the experimental behavior between the two compounds originate from the interplane interactions rather than the intraplane distortion $\alpha$.

\begin{figure}[t]
\begin{center}
\includegraphics[width=18pc]{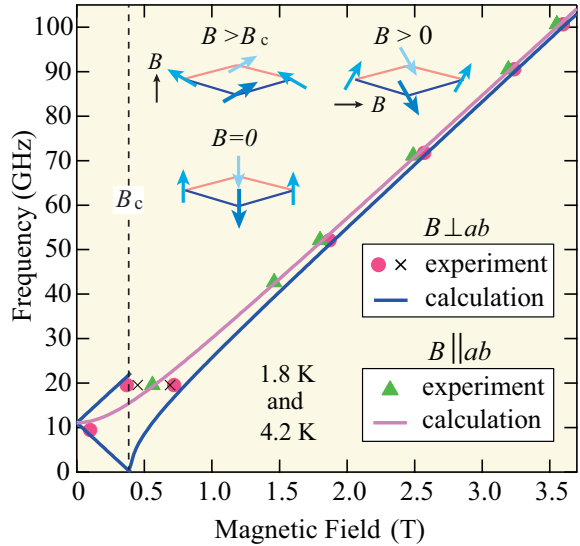}
\caption{(color online) Frequency-field plot of the ESR resonance fields at 1.8 K and 4.2 K in Fig. 7. The solid lines are calculated AF resonance modes with the easy-axis anisotropy. The vertical broken line indicates the spin-flop phase transition at $B_{c}$ for $B{\perp}ab$. The illustrations show schematic views of the spin configurations.}\label{f4}
\end{center}
\end{figure}

\begin{figure}[t]
\begin{center}
\includegraphics[width=18pc]{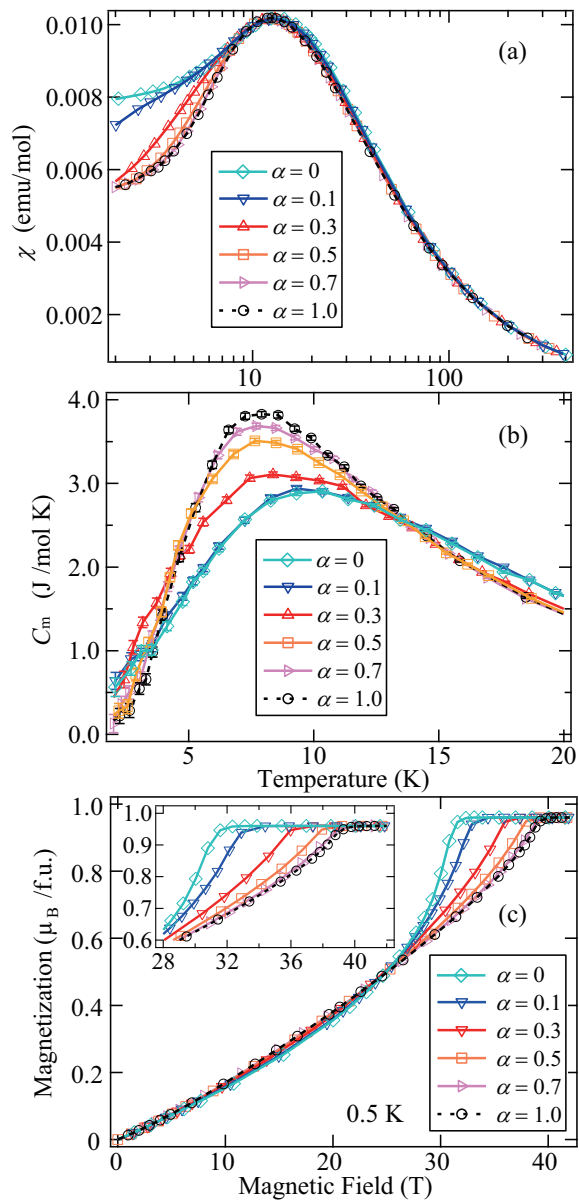}
\caption{(color online) (a) Calculated magnetic susceptibilities, (b) magnetic specific heats, and (c) magnetization curves at 0.5 K for the $S$=1/2 SLHAF with various values of $\alpha=J_{\rm{2}}/J_{\rm{1}}$. The inset shows the expansion of the vicinity of the saturation fields.}\label{f4}
\end{center}
\end{figure}

\subsection{ESR resonance modes}
In this subsection, we evaluate the magnetic anisotropy from the frequency dependence of the ESR resonance signals.
The observed magnetizations indicate the existence of easy-axis anisotropy, which stabilizes the out-of-plane spin structure and induces a spin-flop phase transition, as shown in the schematic picture presented in Fig. 8.
The obtained resonance fields in the frequency-field diagram (Fig. 8) are indeed reminiscent of AF resonance modes with easy-axis anisotropy~\cite{kittel, MnF2}. 
Accordingly, we analyze the experimental results in terms of a mean-field approximation assuming a distorted SLHAF with easy-axis anisotropy. 
Thus, the spin Hamiltonian is expressed as
\begin{equation}
\mathcal {H} = J_{\rm{1}}{\sum^{}_{<ij>}}\textbf{{\textit S}}_i{\cdot}\textbf{{\textit 
S}}_{j}+J_{\rm{2}}{\sum^{}_{<kl>}}\textbf{{\textit S}}_k{\cdot}\textbf{{\textit 
S}}_{l}+D_{\rm{site}}{\sum^{}_{i}}(\mbox{$S$}^{z}_{i})^2-g{\mu _B}{\sum^{}_{i}}\textbf{{\textit S}}_i{\cdot}\textbf{{\textit B}},
\end{equation}
where $D_{\rm{site}}$ is the anisotropy constant of easy-axis type ($D_{\rm{site}}$ $\textless$ 0) and the $z$-axis is perpendicular to the $ab$-plane.
The on-site anisotropy term given in eq.(2) does not involve determination of the ground state in genuine $S$ = 1/2 quantum spin systems.
However, if such $S$ = 1/2 systems exhibit a phase transition to an ordered state, the conventional mean-field approximation becomes effective and describes their ESR resonance modes in the ordered phase~\cite{good_withESR, RbCuCl3, LiCuVO4, LiCu2O2}.
There is no significant difference in the mean-field approximation of the magnetic behavior for two different types of anisotropy: the anisotropic exchange interaction and the on-site anisotropy. 
For both types of anisotropy, the out-of-plane spin structure is stabilized at zero-field, and a spin-flop phase transition occures for $B{\perp}ab$ under the condition that anisotropy is much weaker than exchange coupling.
We adopt the on-site anisotropy as the anisotropic term in the spin Hamiltonian to simplify comparison of the anisotropy energy with that from the dipole field, as described below.    
As the spin structure is described by a two-sublattice model, the free energy $F$ is expressed in the following form, using the mean-field approximation:
\begin{equation}
F=(A+G){\textbf{{\textit M}}_{1}}{\cdot}{\textbf{{\textit M}}_{2}}+K({\mbox{$M$}^{z}_{1}}+{\mbox{$M$}^{z}_{2}})^{2}-({\textbf{{\textit M}}_{1}}+{\textbf{{\textit M}}_{2}}){\cdot}{\textbf{{\textit B}}}, 
\end{equation}
where $A$, $G$, and $K$ are given by
\begin{equation}
A=\frac{2}{N}\frac{2J_1}{(g{\mu _B})^2},\>\> G=\frac{2}{N}\frac{2J_2}{(g{\mu_B})^2},\>\>K=\frac{2}{N}\frac{D_{\rm{site}}}{(g{\mu _B})^2},
\end{equation}
and ${\textbf{{\textit M}}_1}$ and ${\textbf{{\textit M}}_2}$ are the sublattice moments expressed as  
\begin{equation}
{\textbf{{\textit M}}_i} = \frac{N}{2}g{\mu _B}{\textbf{{\textit S}}_i}. 
\end{equation}
Here, $N$ is the number of radicals, and ${\textbf{{\textit S}}_i}$ is the spin on the $i$-th sublattice ($i$=1,2). 
We derive the resonance conditions by solving the equation of motion
\begin{equation} 
{\partial}{\textbf{{\textit M}}_i}/{\partial}t=\gamma[{\textbf{{\textit M}}_i}\times{\textbf{{\textit B}}_i}],
\end{equation}
where $\gamma$ is the gyromagnetic ratio and ${\textbf{{\textit B}}_i}$ is the mean field applied on the $i$-th sublattice moment given by
\begin{equation} 
{\textbf{{\textit B}}_i}=-{\partial}F/{\partial}{\textbf{{\textit M}}_i}.
\end{equation}
To solve the equation of motion, we use a method for the analysis of ABX$_3$-type antiferromagnets~\cite{tanaka}, the efficacy of which has been confirmed for various types of antiferromagnets~\cite{CdCr2O4,CuCrO2,NiGa2S4_JPSJ}.
Assuming precession motion of the sublattice moments around those equilibrium directions, we utilize the following expressions, which represent the motion of the $i$-th sublattice moment:
\begin{equation}
{\textbf{{\textit M}}_i}=({\Delta}M_{i\Acute{x}}\exp(i{\omega}t),{\Delta}M_{i\Acute{y}}\exp(i{\omega}t),|{\textbf{{\textit M}}_i}|),
\end{equation}
where ${\Delta}M_{i\Acute{x}},{\Delta}M_{i\Acute{y}}{\ll}|{\textbf{{\textit M}}_i}|$, and $\Acute{x}$, $\Acute{y}$ and $\Acute{z}$ are the principal axes of the coordinate system on each sublattice moment.
The $\Acute{z}$-axis is defined as being parallel to the direction of each sublattice moment, and the $\Acute{x}$- and $\Acute{y}$-axes are perpendicular to the $\Acute{z}$-axis.

We must consider the spin configuration for each applied field direction in order to obtain the resonance modes ${\omega}$ as functions of $B$.
The spins are aligned along the easy-axis under zero-field conditions, and the discontinuous spin-flop phase transition occurs at $B_{c}$ for $B{\perp}ab$ (see Fig. 8).
The value of $B_{c}$ is expressed as
\begin{equation} 
B_c=\frac{\sqrt{D_{\rm{site}}^{2}-2(J_{1}+J_{2})D_{\rm{site}}}}{g{\mu _B}},
\end{equation}
which corresponds to a zero-field energy gap of resonance modes.
Above $B_{c}$, two sublattices are tilted from the 2D plane with equivalent angles (see Fig. 8).
For $B{\parallel}ab$, where the external field is applied perpendicular to the easy-axis, two sublattices are tilted from the easy-axis with equivalent angles (see Fig. 8).
The angles between the sublattice moment and the external field for both directions are determined by minimizing the free energy. 
Further, $D_{\rm{site}}/k_{\rm{B}}=-0.0055$ K is obtained from eq.(9), using $B_{c}$ = 0.4 T, $J_{\rm{1}}/k_{\rm{B}}$ = 16.5 K, and  $J_{\rm{2}}/k_{\rm{B}}$ = 9.2 K ($\alpha= 0.56$).
Then, the ${\omega}$ values are obtained by solving eq.(6) numerically.
The calculated results demonstrate typical AF resonance modes with easy-axis anisotropy in a two-sublattice model~\cite{kittel, MnF2} exactly, and we obtain good agreement between experiment and calculation, as shown in Fig. 8. 
Although some field-independent ${\omega}$ are expected, only field-dependent ${\omega}$ detectable via our field-sweep measurements are displayed here.
For $B{\perp}ab$, two gapped modes appear below $B_{c}$.
The lower-frequency mode becomes soft and exhibits a discontinuous change at $B_{c}$.
For $B{\parallel}ab$, one gapped mode exhibits a gradual increase with increasing field.

The mean-field approximation with anisotropic exchange interactions gives the same AF resonance modes as those obtained with on-site anisotropy. 
In that case, the spin Hamiltonian is written as,
\begin{equation}
\begin{split}
\mathcal {H} = J_{\rm{1}}{\sum^{}_{<ij>}}(\mbox{$S$}^{x}_{i}\mbox{$S$}^{x}_{j}+\mbox{$S$}^{y}_{i}\mbox{$S$}^{y}_{j}+\delta \mbox{$S$}^{z}_{i}\mbox{$S$}^{z}_{j})+\\J_{\rm{2}}{\sum^{}_{<kl>}}(\mbox{$S$}^{x}_{k}\mbox{$S$}^{x}_{l}+\mbox{$S$}^{y}_{k}\mbox{$S$}^{y}_{l}+\delta \mbox{$S$}^{z}_{k}\mbox{$S$}^{z}_{l})-g{\mu _B}{\sum^{}_{i}}\textbf{{\textit S}}_i{\cdot}\textbf{{\textit B}},
\end{split}
\end{equation}
where $\delta$ is the anisotropy constant for the $z$-components ($\delta$ $>$ 1), and the $z$-axis is perpendicular to the $ab$-plane.
The spin configuration for each applied field direction is identical to those in the on-site case, and $B_{c}$ is given by
\begin{equation} 
B_c=\frac{\sqrt{\delta^{2}-1}(J_{\rm{1}}+J_{\rm{2}})}{g{\mu _B}}
\end{equation}
A value of $\delta=1.00021$ is obtained using $B_{c}$ = 0.4 T, $J_{\rm{1}}/k_{\rm{B}}$ = 16.5 K, and  $J_{\rm{2}}/k_{\rm{B}}$ = 9.2 K ($\alpha= 0.56$), and we obtain almost the same resonance modes as those in Fig. 8 by solving the equation of motion corresponding to eq.(6) numerically.

\subsection{Dipole field anisotropy}
The magnetic anisotropy in organic radical systems is known to be quite small and almost isotropic at experimental temperatures.
In the present material, the energy scale of the magnetic anisotropy is also evaluated to be small from the ESR analysis.
Here, we consider dipole-dipole interactions as a possible origin of the anisotropy.
The anisotropic field arising from the dipole-dipole interactions can be calculated via a method used in the case of CuCl$_{2}{\cdot}$2H$_2$O~\cite{dipole}. 
Assuming the collinear spin configuration at zero-field described by the two sublattice model, we regard each lattice point as having ${\textbf{{\textit m}}_1}$ or ${\textbf{{\textit m}}_2}$, which are oppositely oriented spin magnetic moments.
Then, each component of the magnetic dipole field ${\textbf{{\textit B}}_1}$ at a lattice point with ${\textbf{{\textit m}}_1}$ is produced by the spins on the other lattice points and given by 
\begin{equation} 
\begin{split}
\mbox{$B$}^{x}_{1}={\varPhi}_{a1}\mbox{$m$}^{x}_{1}+{\varPhi}_{a2}\mbox{$m$}^{x}_{2},\\
\mbox{$B$}^{y}_{1}={\varPhi}_{b1}\mbox{$m$}^{y}_{1}+{\varPhi}_{b2}\mbox{$m$}^{y}_{2},\\
\mbox{$B$}^{z}_{1}={\varPhi}_{c1}\mbox{$m$}^{z}_{1}+{\varPhi}_{c2}\mbox{$m$}^{z}_{2},
\end{split}
\end{equation}
where the $x$-, $y$-, and $z$-axes are defined as being parallel to the $a$- and $b$-axes and perpendicular to the $ab$-plane, respectively.
${\textbf{{\textit B}}_2}$ is also derived through appropriate permutation of the moments.
${\varPhi}_{ai}$, ${\varPhi}_{bi}$, and ${\varPhi}_{ci}$ are the dipole sums and expressed as
\begin{equation} 
\begin{split}
{\varPhi}_{ai}=-{\Sigma}_{j}^{(i)}[1-3(\frac{r_{1j}^{x}}{r_{1j}})^{2}]r_{1j}^{-3},\\
{\varPhi}_{bi}=-{\Sigma}_{j}^{(i)}[1-3(\frac{r_{1j}^{y}}{r_{1j}})^{2}]r_{1j}^{-3},\\
{\varPhi}_{ci}=-{\Sigma}_{j}^{(i)}[1-3(\frac{r_{1j}^{z}}{r_{1j}})^{2}]r_{1j}^{-3},
\end{split}
\end{equation}
where ${\Sigma}_{j}^{(i)}$ is taken over all the distances $r_{1j}$ between one lattice point with ${\textbf{{\textit m}}_1}$ and the neighboring lattice points $j$ with ${\textbf{{\textit m}}_i}$.
We obtain a sufficient condition for convergence considering the lattice points up to 120 sites.  
The anisotropy energy arising from the dipole-dipole interactions $F_{\rm{dip}}=-(\textbf{{\textit m}}_1{\cdot}\textbf{{\textit B}}_{1}+\textbf{{\textit m}}_2{\cdot}\textbf{{\textit B}}_{2})$ is expressed as   
\begin{equation} 
\begin{split}
F_{\rm{dip}}=({\varPhi}_{a2}-{\varPhi}_{a1})[(\mbox{$m$}^{x}_{1})^2+(\mbox{$m$}^{x}_{2})^2]\\
+({\varPhi}_{b2}-{\varPhi}_{b1})[(\mbox{$m$}^{y}_{1})^2+(\mbox{$m$}^{y}_{2})^2]\\
+({\varPhi}_{c2}-{\varPhi}_{c1})[(\mbox{$m$}^{z}_{1})^2+(\mbox{$m$}^{z}_{2})^2].
\end{split}
\end{equation}
Accordingly, we can express the anisotropy energy as 
\begin{equation} 
F_{\rm{dip}}=D_{\rm{dip}}{\sum^{}_{i}}(\mbox{$S$}^{z}_{i})^2+E_{\rm{dip}}{\sum^{}_{i}}[(\mbox{$S$}^{x}_{i})^2-(\mbox{$S$}^{y}_{i})^2]+A,
\end{equation}
where $A$ is a constant independent of the moment direction. 
The on-site anisotropy constants per site $D_{\rm{dip}}$ and $E_{\rm{dip}}$ are expressed by 
\begin{equation} 
\begin{split}
D_{\rm{dip}}=\frac{2}{3}({\varPhi}_{c2}-{\varPhi}_{c1})(g{\mu _B})^2,\\
E_{\rm{dip}}=\frac{1}{2}[({\varPhi}_{a2}-{\varPhi}_{a1})-({\varPhi}_{b2}-{\varPhi}_{b1})](g{\mu _B})^2.
\end{split}
\end{equation}
To calculate the dipole sums, we assume that each spin is almost localized at the center of the verdazyl ring that contains four N atoms, and we use the average position of the N atoms as the lattice point.
Substituting the calculated dipole sums into eq. (16), we obtain $D_{\rm{dip}}/k_{\rm{B}}$=-0.030 K and $E_{\rm{dip}}/k_{\rm{B}}$=-0.013 K.
The calculated results indicate the presence of the easy-axis perpendicular to the $ab$-plane ($z$-axis), which is consistent with the experimental evaluations.

Here, we discuss the cause of the difference in the value of the easy-axis anisotropy constant between the calculation and the evaluation via experiment.
To calculate the dipole sums exactly, it is necessary to consider the spin distribution on the molecule~\cite{dipole_radical}.
However, we assumed localized spin in our calculation because exact consideration of the spin distribution in this molecule generates unnecessary difficulties in the calculation due to the large atom numbers in the chemical formula.   
Although it has been confirmed that the localized spin model is effective for describing magnetic behavior in verdazyl-based compounds~\cite{2Cl6FV,3Cl4FV, pBrV,3ladders, okabe, Zn}, calculation of the dipole field via this approximation may induce some differences in such small anisotropy values.
As a major factor in this discrepancy, we must consider the fact that our calculations for the dipole field are based on an ordered moment size.
The quantum fluctuations are expected to reduce the magnetic moment by approximately 40 ${\%}$ for an $S$=1/2 SLHAF~\cite{square}.
The nonlinear increase of the magnetization curve observed in the ordered phase actually demonstrates the magnetic moment reduction due to the quantum fluctuations in this compound.
Thus, if we assume reduction of the local moment, the anisotropy energy $F_{\rm{dip}}$, which decreases with the square of the moment size, should be substantially reduced.
For instance, considering 40 ${\%}$ reduction of the magnetic moment, it is required to multiply the right side of Eq. (16) by 0.6$^2$. 
Therefore, we confirm that the reduced magnetic moment is the main cause of the difference between the calculated and actual values of the anisotropy constant.


\section{Summary}
We have succeeded in synthesizing single crystals of the verdazyl radical $\alpha$-2,3,5-Cl$_3$-V.
$Ab$ $initio$ MO calculations indicate that the two dominant AF interactions, $J_{\rm{1}}$ and $J_{\rm{2}}$ ($\alpha =J_{\rm{2}}/J_{\rm{1}}\simeq 0.56$), form $S$ = 1/2 distorted square lattices partially connected by the weak AF interaction $J_{\rm{3}}$.
We performed magnetization, specific heat, and multi-frequency ESR measurements on the single crystals. 
The magnetic susceptibility and the magnetization curve were explained based on the $S$ = 1/2 SLHAF using the QMC method, and the effect of the lattice distortion and the interplane interaction contribution were clarified. 
It was confirmed that the chain-like lattice distortion of the present model is approximately $0.5<\alpha<1$, which is sufficiently small to avoid affecting the intrinsic behavior of the uniform $S$ = 1/2 SLHAF.
Furthermore, it is considered that the two-dimensionality should be enhanced in the present lattice owing to the smallest interplane coordination number.
In the low-temperature regions, we observed a phase transition to the AF ordered state at approximately $T_{\rm{N}}$ = 6.4 K for zero-field conditions and anisotropic magnetic behavior for $T{\textless}T_{\rm{N}} $.
The ESR resonance signals elucidated the corresponding critical and anisotropic behavior in the temperature dependence of the resonance field and the linewidth.
The frequency dependence of the ESR resonance fields for $T{\textless}T_{\rm{N}} $ was well explained by a mean-field theory considering the out-of-plane easy-axis anisotropy, which causes a spin-flop phase transition at $B_{\rm{c}}$=0.4 T for $B{\perp}ab$. 
The anisotropic energy derived from the dipole-dipole interactions actually indicates the presence of the easy-axis perpendicular to the $ab$-plane.
These results demonstrate that $\alpha$-2,3,5-Cl$_3$-V is a new model compound with an $S$ = 1/2 square-based lattice.
The lattice distortion, anisotropy, and interplane interaction in this model should have perturbative effects on the ground state of the $S$ = 1/2 SLHAF, and further investigations will yield quantitative information regarding the related dynamical properties.


\begin{acknowledgments}
This research was partly supported by Grant for Basic Science Research Projects from KAKENHI (No. 15H03695, No. 15K05171, and No. 17H04850), the CASIO Science Promotion Foundation. A part of this work was carried out at the Center for Advanced High Magnetic Field Science in Osaka University under the Visiting Researcher's Program of the Institute for Solid State Physics, the University of Tokyo and the Institute for Molecular Science.
\end{acknowledgments}


\end{document}